\documentclass{sig-alternate}
\usepackage[latin9]{inputenc}
\usepackage{textcomp}
\usepackage{amstext}
\usepackage{amssymb}
\usepackage{graphicx}
\begin{document}
\conferenceinfo{KDD}{'14 El Paso, Texas USA}
\title{Overlapping Community Detection Optimization \\ and Nash Equilibrium }
\subtitle{[Overlapping Community Detection]
}

%%%%%%%%%%%%%%%%%%%%%%%%%%%%%% LyX specific LaTeX commands.
\numberofauthors{2} %  in this sample file, there are a *total*

\author{
%% Because html converters don't know tabularnewline
\alignauthor
Michel Crampes\\
       \affaddr{LGI2P Laboratory}\\
       \affaddr{Ecole des Mines, Parc Georges Besse}\\
       \affaddr{30035, Nîmes, France}\\
       \email{michel.crampes@mines-ales.fr}
\alignauthor
Michel Plantié\\
       \affaddr{LGI2P Laboratory}\\
       \affaddr{Ecole des Mines, Parc Georges Besse}\\
       \affaddr{30035, Nîmes, France}\\
       \email{michel.plantie@mines-ales.fr}
}
\maketitle

\begin{abstract}
Community detection using both graphs and social networks is the focus of many algorithms. Recent methods aimed at optimizing the so-called modularity function proceed by maximizing relations within communities while minimizing inter-community relations. 

However, given the NP-completeness of the problem, these algorithms are heuristics that do not guarantee an optimum. In this paper, we introduce a new algorithm along with a function that takes an approximate solution and modifies it in order to reach an optimum. This reassignment function is considered a 'potential function' and becomes a necessary condition to asserting that the computed optimum is indeed a Nash Equilibrium. We also use this function to simultaneously show partitioning and overlapping communities, two detection and visualization modes of great value in revealing interesting features of a social network. Our approach is successfully illustrated through several experiments on either real unipartite, multipartite or directed graphs of medium and large-sized datasets.

\end{abstract}

%\keywords{Overlaping Communities, Community Detection, Nash equilibrium%}
%\maketitlepage


\section{Introduction}

%\vspace{-0.2cm}
 With social networks  being so widespread on the Internet, community
detection in social graphs has recently become a major field of research.
Many algorithms have been proposed (see several surveys on
this topic in \cite{Papadopoulos2011,Yang2010,Porter2009} and a
more detailed survey in \cite{Fortunato2009}). 
Most of these algorithms take unipartite graphs as inputs and produce partitioned
communities.
In the real world however, individuals are present in several communities. Overlapping
is a common characteristic. Recent works have focused on detecting overlapping
communities, mainly on unipartite graphs. 
In a previous article we focused on extracting overlapping communities in 
bipartite graphs through the use of Galois Lattices \cite{Crampes2012}. 
In a more recent paper, we proposed a different method \cite{Crampes2013} that
 uses any partitioning algorithm for unipartite graphs in order to process any type of graph
(i.e. Unipartite, Bipartite, Oriented) and then extract partitioned and overlapping
communities. We have actually  made use the so-called Louvain Algorithm \cite{Blondel2008}.
Applied to well-known benchmarks we have obtained results as good  or better than any other author.
More specifically, we have been able to represent both partitioned and overlapping communities. 
However, despite its efficiency, our method based on the Louvain algorithm contains some important limitations, which we would like to address in this paper. Like any other heuristic, ours yields partitioned communities, 
in which the distribution of vertices is not guaranteed to be a global modularity optimum. In particular, potential overlapping clearly shows the lack of an optimum. To more closely approximate the optimum, we introduce a reassignment function that allows taking into account the wishes of each vertex to change communities. This function is then complemented by a stability condition, which is actually a Nash Equilibrium. The results are greatly enhanced. Convergence to a Nash Equilibrium provides not only a way to reach a local optimum within an acceptable computational time, but also a powerful means for semantic interpretation of agent assignment and reassignment to communities. Our paper concludes with an open perspective on these results.

\section{State-of-the-art }

 As stated above, several state-of-the-art assessments have already addressed the community detection problem; these mainly focus on unipartite graph partitioning. The calculations performed are based on maximizing a mathematical criterion, in most cases modularity (introduced by Newman \cite{Newman2004})
 in representing the maximum number of connections within each community and a minimum number of links with external communities. Various methods have been developed to identify the optimum, e.g. greedy algorithms \cite{Newman2004a,Noack2008},
spectral analyses \cite{Newman2006}, or a search for the most centric
edges \cite{Newman2004}. One of the most efficient greedy algorithm
for extracting partitioned communities from large (and possibly weighted)
graphs is Louvain \cite{Blondel2008}. In a very comprehensive state-of-the-art
report \cite{Fortunato2009} other new partitioned community detection
methods are described. 
Several authors have since extended modularity to bipartite graphs, 
first of all by adapting the formulae \cite{Murata2010,Suzuki2009}.  
Barber \cite{Barber2007} however derived a modularity expression for bipartite graphs from Newman's
 that was subsequently used by several authors to apply classical methods, such as Simulated Annealing 
 \cite{Guimera2007}, spectral clustering \cite{Barber2007},
genetic algorithms \cite{Nicosia2009}, Label propagation  \cite{LiuXin2010}, 
or spectral dichotomic analysis\cite{Leicht2007}. 
Overlapping community detection strategies from unipartite graphs are very often
 simply extensions of partitioned extraction methods. Reference 
\cite{Palla2005a} uses k-clique percolation methods, while \cite{Davis2008} uses
a random march in a graph and \cite{Gregory2009} designs label propagation algorithms.
Some authors propose specific methods: \cite{Wu2012a} extract overlapping in partitioned communities, 
\cite{Reichardt2006a} combine the Pott's spin interaction model with  Simulated Annealing,
 \cite{Lancichinetti2009} optimizes a local statistic function and \cite{Evans2009} 
treats a dual problem involving a partitioning of weighted links. It is rare to find research work focusing on both bipartite graphs and overlapping communities. We have identified some method extensions, like overlapping bi-clique extraction \cite{SuneLehmannMartinSchwartzLarsKaiHansen2008}.
Other original methods use well-known results in Galois Lattices (\cite{Crampes2012} and \cite{Roth2008}),
 though their algorithms and representations
remain complex. In a recent article \cite{Crampes2013}, 
we demonstrated the possibility of unifying unipartite, bipartite and oriented graphs, 
in addition to extracting both overlapping and partitioned communities; 
the overlapping properties are computed using a simple simultaneous membership function. 
We decided to apply the well-known Louvain Algorithm \cite{Blondel2008},
which iteratively aggregates the graph vertices in order to maximize the modularity function. In the general case however, this algorithm, like any heuristic, merely produces an approximate result, i.e. the algorithm stops once modularity can increase no further. Moreover, our overlapping function shows that some vertices may have an incentive to belong to other communities and that such changes may increase or decrease the results already achieved, meaning that the present result is not stable. The research on stable communities has not been extensively studied, as authors tend to be satisfied with their own results on their own algorithms, with comparisons to other authors' results. Yet the body of research on network stability from the perspective of "game theory" is quite prolific and has provided numerous scientific articles, e.g. \cite{Nisan2007}.
According to this approach, attaining the stability of n agents, all of whom choose their own strategies with the help of satisfaction functions, presumes that a Nash Equilibrium actually exists. When applied to community detection, the problem can be stated as follows: find the conditions yielding a Nash Equilibrium such that no vertex would wish to leave the community where it has been assigned. To the best of our knowledge, Nash Equilibrium has seldom been used to detect communities, though 
 \cite{RNarayanam2012} applied it to unipartite graphs.  These authors used vertex connectivity to reach a Nash Equilibrium, 
yet without measuring the resulting modularity. 
\cite{Chen2011} focused on unipartite graphs to extract overlapping communities with Nash Equilibrium as the lone guiding objective. 
Yet the experimental results do not seem to relate to published material on this subject. In this article, 
we draw a distinction with these authors by first seeking an approximate robust solution with the Louvain algorithm and then performing reallocations in order to converge towards a Nash Equilibrium. Moreover, we are able to apply our approach to all three types of graphs.

\section{Partitioned Community \\Detection and Overlap}

%\vspace{-0.2cm}

\subsection{Modularity}

%\vspace{-0.2cm}

We have shown in \cite{Crampes2013} that detecting partitioned communities in unipartite, bipartite or directed graphs may be considered as community detection in unipartite graphs. Consequently, we introduced definitions and proved all properties on unipartite graphs, before applying them to the unipartite and bipartite graphs in our experiments. The common consensus among authors in the community detection discipline is to reach an approximate result maximizing Newman's modularity \cite{Newman2006}.
This formula measures the quality of graph partition. Let's define a unipartite graph 
 $G=(N,E)$ represented with its adjacency matrix $A$ ,
the modularity $Q$ of a graph partition is defined as:
\begin{eqnarray}
Q=\frac{1}{2m}\sum_{i,j}\left[A{}_{ij}-\frac{k_{i}k_{j}}{2m}\right]\delta(c_{i},c_{j})
 \label{equ1}
\end{eqnarray}

where $A{}_{ij}$ is the weight of the link between $i$ and $j$,
$k_{i}=\sum_{j}A{}_{ij}$ is the sum of weights of edges attached to vertex $i$, 
$c_{i}$ is the community of $i$, the Kronecker function $\delta(u,v)$ is equal to $1$
if $u=v$ and $0$ elsewhere, and finally $m=1/2\sum_{ij}A{}_{ij}$.
We are only considering binary graphs herein, in which case weights  $A{}_{ij}$ have just two values: 
 $1$ or $0$, depending on whether the edge exists or not.  We can interpret this formula as follows: for all communities, modularity is the weighted sum of the differences between links inside the community (term $A{}_{ij}$) 
and the probability of these links (term $\frac{k_{i}k_{j}}{2m}$ whose numerator is the product
of  margins corresponding to cell $i,j$).
Applying this function with many types of algorithms yields satisfactory results. For example, it is possible to find communities
of \textit{ad-hoc} constructed graphs.
Fortunato \cite{Fortunato2009} however demonstrated that this function tends to merge small communities and,
 consequently, swallows small granularities. 
For bipartite graphs (and directed graphs, which can be reduced to bipartite graphs, see \cite{Crampes2013}),
other formulations have been offered. In particular, Barber's formulation \cite{Barber2009}
has won consensus among researchers: it is not very different from Newman's. 
Nonetheless, we will still use Newman's formulation regardless of the graph type 
considered in order to remain consistent with our previous works published, in unifying the three types of graphs.

\subsection{Bipartite graph partitioning \label{sub:Modularity-and-bimodularity}}

This section will show how to transform a bipartite graph into a unipartite graph,
 in order to apply all algorithms of other graph types, and then explain the consequences of this action.

\subsubsection{Turning bipartite graphs into unipartite graphs}

In formal terms, a bipartite graph $G=(U,V,E)$ is a graph $G'=(N,E)$
where node set $N$ is the union of two independent sets $U$ and
$V$ and moreover the edges only connect pairs of vertices $(u,v)$
where $u$ belongs to $U$ and $v$ belongs to $V$. 

$N=U\cup V$, $U\cap V=\emptyset$, $E\subseteq U\times V$.

Let: $r=|U|$ and $s=|V|,$ then$|N|=n=r+s$

The non-weighted biadjacency matrix of a bipartite graph $G=(U,V,E)$
is a $r\times s$ matrix $B$ in which:

$B_{i,j}=1\: iff(u_{i},v_{j})\in E$ and $B_{i,j}=0\: iff(u_{i},v_{j})\notin E$. 

Let's point out that the row margins in $B$ represent the
degrees of nodes $u_{i}$ while the column margins
represent the degrees of nodes $v_{j}$. Conversely, in $B^{t}$,
the transpose of $B$, row margins represent the degrees of nodes
$v_{j}$ and column margins represent the degrees
of nodes $u_{i}$. Let's now define the off-diagonal block square
matrix $A'$ :

$A'=\left(\begin{array}{cc}
0_{r} & B\\
B^{t} & 0_{s}
\end{array}\right)$ where $0_{r}$ is an all zero square matrix of order $r$ and $0_{s}$
is an all zero square matrix of order $s$. 

This symmetric matrix is the adjacency matrix of the unipartite graph
$G'$ where nodes types have not been distinguished. It is possible
to apply any algorithm to $G'$ for the purpose of extracting communities
from unipartite graphs. $A'$ is also the off-diagonal adjacency matrix
of bipartite graph $G$. Consequently the communities 
detected in $G'$ are also detected in $G$.
The question then is to determine the validity 
of this secondary result: what is the quality of partitioning for  $G$ when applying
an unipartite graph partitioning algorithm on $G'$? Barber
\cite{Barber2007} and Liu/Murata \cite{LiuXin2010} also introduced the block matrix 
as a means of detecting communities in bipartite graphs. 
We notice below however that these authors have not taken all consequences of this approach into account.

\subsubsection{Extending modularity to bipartite graphs }

Modularity is an indicator often used to measure the quality of graph
partitions \cite{Newman2004}. First defined for unipartite graphs,
 several modularity variants have been proposed for bipartite graph partitioning as well as for overlapping communities. 
More recently, several authors introduced modularity into bipartite graphs using a probabilistic analogy with the modularity for unipartite graphs.
 Yet when applying unipartite graph modularity optimization algorithms to bipartite graphs, another expression of probabilistic modularity emerges, as will be presented hereafter. 

Let $G=(U,V,E)$ be a bipartite graph with its biadjacency matrix
$B$ and the unipartite graph $G'$ with the adjacency off-diagonal
block matrix $A'$. Let's also consider Newman's modularity \cite{Newman2004}
for this graph $G'$: it is a function $Q$ of both matrix $A'$ and
the communities detected in $G'$:

\begin{equation}
Q=\frac{1}{2m}\sum_{i,j}\left[A'_{ij}-\frac{k_{i}k_{j}}{2m}\right]\delta(c_{i},c_{j})\label{eq:1}
\end{equation}

where $A'_{ij}$ represents the weight of the edge between $i$ and
$j$, $k_{i}=\sum_{j}A'_{ij}$ is the sum of weights for the edges attached to vertex
 $i$, $c_{i}$ denotes the community to which vertex
$i$ is assigned, the Kronecker function $\delta(u,v)$
equals $1$ if $u=v$ and $0$ otherwise and $m=1/2\sum_{ij}A'_{ij}$.
Hereafter we only consider binary graphs and weights are equal to
$0$ or $1$.

After several transformations we show  (for further details, see \cite{Crampes2013})
that this modularity can also be written using the biadjacency matrix
$B$ of the bipartite graph $G=(U,V,E)$:

\begin{equation}
Q^{B}=\frac{1}{m}\sum_{ij}[B_{ij}-\frac{(k_{i}+k_{j})^2}{4m}]\delta(c_{i},c_{j})\label{eq:9}
\end{equation}

where $k_{i}$ is the margin of row $i$ in $B$, $k_{j}$ the margin
of column $j$ in $B$ and $m=\sum_{ij}B_{ij}=\frac{1}{2}\times\sum_{ij}A'_{ij}=m$
in (\ref{eq:1}).

Another interesting formulation to be introduced is as follows (\cite{Crampes2013}): 

\begin{equation}
Q^{B}=\sum_{c}[\frac{|e_{c}|}{m}\text{\textendash}(\frac{(d_{u|c}+d_{v|c})}{2\times m})^2]\label{eq:3-2}
\end{equation}

where $|e_{c}|$ is the number of edges in community $c$, and $d_{w|c}$
is the degree of node $w$ belonging to $c$.

This modularity formulation is the same as Newman's, 
yet with more detailed information: it explicitly shows 
that both sets of nodes are structurally associated in the same communities. 

Since in the general case $B$ is asymmetric, this definition thus characterizes modularity 
for bipartite graphs after their extension into unipartite graphs. 
It then becomes possible to apply any partitioning algorithm for unipartite graphs  
to matrix $A'$ and obtain a result where both types of nodes are bound in the same communities, 
except in the case of singletons (i.e. nodes without edges). 
This definition from unipartite graph modularity, in recognizing its ability to bind both types of nodes, 
has been compared in  \cite{Crampes2013}
with other authors' modularity models for bipartite graphs.

\subsection{Partitioned Community Detection}

Among the numerous algorithms proposed by researchers to extract partitioned communities, the Louvain Algorithm \cite{Blondel2008}
is remarkable for its efficiency and quality of results; it uses a hierarchical method to build communities by finding at each step the modularity optimum for each vertex. At each step, communities from the previous steps are replaced by a vertex representing all its components. Consequently, the graph is progressively reduced to reach a modularity optimum.

\subsection{Overlapping Community Detection \\and Analysis \label{sec4}}

Several approaches exist to detect overlapping among communities. In scientific corpora dedicated to the detection of overlapping communities, vertices are associated with several communities. Such is the case for example in \cite{Palla2005a} and \cite{Chen2011}.
In our case, we start by identifying partitioned communities; then, an overlapping function is used to detect instabilities in community membership. We are able to simultaneously obtain partitioned communities and their overlap. Several overlapping functions may be applied; some only detect overlapping, while others like ours measure the overlapping ratio. Our function, called "legitimacy", offers the advantage of detailing for each vertex its membership degree to each community. For each vertex $u_{i}$, the membership degree (or membership legitimacy) to the c community is measured by the number of edges related to community vertices divided by the community size:
\begin{eqnarray}
L(u_{i}\in c)=\frac{\sum{}_{j}A_{ij}\delta(c_{j})}{|\{v\in c\}|}
\end{eqnarray}

This function may be interpreted as follows: the more a vertex is attracted by a community, the higher its relative number of relations with this community, independent of its size. This function is both intuitive and simple. For example, for the SW benchmark presented in Section \ref{sub:SW}
we can observe that $c_{1}$ contains 7 events, $c_{2}$ 5 events and $c_{3}$ 2 events.
For $w_{1}$, we obtain a legitimacy value of $\frac{2}{7}$ for $c_{1}$, $\frac{1}{5}$ for $c_{2}$ and $\frac{1}{2}$
for $c_{3}$. Despite its simplicity, this function shows that some vertices may belong to several communities and that these communities may be unstable with suboptimal modularity. The Louvain algorithm, like any other "glutton" type algorithm, yields an approximate solution. To increase overall modularity, we have introduced a new reassignment function, which will be the topic of the following section.
\newline
 
 \section{Reassignment}

\subsection{Reassignment Modularity (RM) function} \label{RM}

To define this Reassignment function, let's introduce a variant of Equation \ref{equ1}.
Let $C_{i}$ be a community, $|e_{i}|$ the number of edges in  $C_{i}$ and $d_{C_{i}}$ 
the sum of degrees of each node in $C_{i}$. 
Then $Q=\sum_{i}\left[\frac{|e_{i}|}{m}-\frac{(d_{C_{1}})^{2}}{(2m)^{2}}\right]$.
Reassigning a vertex $w$ from $C_{1}$ to $C_{2}$ increases or decreases modularity.
This variation is defined as the Modularity Reassignment measure
 $RM_{w:C_{1}\rightarrow C_{2}}$ =$Q_{w\in C_{2}}$-$Q_{w\in C_{1}}$
where $Q_{w\in C_{k}}$ is the modularity value for $w\in C_{k}$
and $C_{1}\neq C_{2}$.
Let $l_{w|i}$ be the number of edges between a vertex $w$ and all other
vertices $w'$, such as $w'\in C_{i}$, 	and let $d_{w}$ be the degree of $w$.
It is now considered that vertex $w$ belonging to $C_{1}$ is removed from this community
and then reassigned to another community $C_{2}$. Then

$Q_{w\in C_{1}}$ = $[\frac{1}{m}|e_{1}|-\frac{(d_{C_{1}})^{2}}{(2m)^{2}}+\frac{1}{m}|e_{2}|-(\frac{(d_{C_{2}})^{2}}{(2m)^{2}})]$
+ $K_{others}$ where $K_{others}$ is the modularity contribution
of communities other than $C_{1}$ and $C_{2}$. $K_{others}$
does not change when  reassigning a vertex from $C_{1}$ to $C_{2}$.

$Q_{w\in C_{2}}$ = $[\frac{1}{m}(|e_{1}|-l_{w|1})+\frac{1}{m}(|e_{2}|+l_{w|2})-(\frac{(d_{C_{1}}-d_{w})^{2}}{(2m)^{2}}+\frac{(d_{C_{2}}+d_{w})^{2}}{(2m)^{2}})]$
+ $K_{others}$

then $RM_{w:C_{1}\rightarrow C_{2}}=Q_{w\in C_{2}}$-$Q_{w\in C_{1}}$
= $[\frac{1}{m}(|e_{1}|-l_{w|1})+\frac{1}{m}(|e_{2}|+l_{w|2})-(\frac{(d_{C_{1}}-d_{w})^{2}}{(2m)^{2}}+\frac{(d_{C_{2}}+d_{w})^{2}}{(2m)^{2}})]-[\frac{1}{m}|e_{1}|-\frac{(d_{C_{1}})^{2}}{(2m)^{2}}+\frac{1}{m}|e_{2}|-(\frac{(d_{C_{2}})^{2}}{(2m)^{2}})]$
and after simplifying:

\begin{eqnarray}
RM_{w:C_{1}\rightarrow C_{2}}=
\frac{l_{w|2}-l_{w|1}}{m}-\frac{[d_{w}^{2}+d_{w}(d_{C_{2}}-d_{C_{1}})]}{2m{}^{2}}\label{eq:11}
\end{eqnarray}

When applying Equation \ref{eq:11}, it is possible to verify that if we reassign
$w$ from $C_{1}$ to $C_{2}$ and then again from
$C_{2}$ to $C_{1}$, modularity does not change (this  calculation may be verified
in different ways) and $RM_{w:C_{1}\rightarrow C_{2}\rightarrow C_{1}}=0$.
Accordingly: $RM_{w:C_{1}\rightarrow C_{2}}=-RM_{w:C_{2}\rightarrow C_{1}}$.

\subsection{Effects of Reassigning vertices}

Next, we assume that a first pass reassignment calculation has been performed, following which 
$w$, a graph vertex is moved from $C_{1}$ to $C_{2}$, because of a positive reassignment value RM.
Let $z$ be another graph vertex: we can observe the reassignment value for this vertex after moving $w$.
The difference in reassignment measure for vertex $z$ from the previous step to the present step
is computed as follows:
we seek $RM_{z:C_{from}\rightarrow C_{to}}^{1}-RM_{z:C_{from}\rightarrow C_{to}}^{0}$
where $RM_{z:C_{from}\rightarrow C_{to}}^{0}$ is the reassignment measure
of $z$ before the move of $w$ and $RM_{z:C_{from}\rightarrow C_{to}}^{1}$
its new value afterwards. Depending on the values of $C_{from}$ and $C_{to}$ we find
the following results: Let $\triangle R_{z}=[\{w,z\}-\frac{1}{(2m)}d_{z}d_{w}]\frac{1}{m}$,
in which $\{w,z\}$ represents the edge between $w$ and $z$. In the absence of an edge, this value is equal to zero.

The following table lists the various individually-computed rectifications of the $z$ vertex reassignment value. 
$C_{3}$ and $C_{4}$ are communities different from $C_{1}$ and $C_{2}$.
\newline

\begin{center}
\begin{tabular}{|c|c|c|c|c|}
\hline 
to\textbackslash{}from  & $C_{1}$  & $C_{2}$  & $C_{3}$  & $C_{4}$\tabularnewline
\hline 
\hline 
$C_{1}$  & 0  & -2$\triangle R_{z}$  & -$\triangle R_{z}$  & -$\triangle R_{z}$\tabularnewline
\hline 
$C_{2}$  & 2$\triangle R_{z}$  & 0  & $\triangle R_{z}$  & $\triangle R_{z}$\tabularnewline
\hline 
$C_{3}$  & $\triangle R_{z}$  & -$\triangle R_{z}$  & 0  & 0\tabularnewline
\hline 
$C_{4}$  & $\triangle R_{z}$  & -$\triangle R_{z}$  & 0  & 0\tabularnewline
\hline 
\end{tabular}
\end{center}

We may infer the following interesting properties:

$(RM_{z:C_{1}\rightarrow C_{2}}^{1}-RM_{z:C_{1}\rightarrow C_{2}}^{0})=$

$(RM_{z:C_{1}\rightarrow C_{3}}^{1}-RM_{z:C_{1}\rightarrow C_{3}}^{0})+(RM_{z:C_{3}\rightarrow C_{2}}^{1}-RM_{z:C_{3}\rightarrow C_{2}}^{0})$

= $2\triangle R_{z}$ et $(RM_{z:C_{2}\rightarrow C_{1}}^{1}-RM_{z:C_{2}\rightarrow C_{1}}^{0})=$

$(RM_{z:C_{2}\rightarrow C_{3}}^{1}-RM_{z:C_{2}\rightarrow C_{3}}^{0})$
$+(RM_{z:C_{3}\rightarrow C_{1}}^{1}-RM_{z:C_{3}\rightarrow C_{1}}^{0})$

= $-2\triangle R_{z}$

We note that the matrix is antisymmetric, 

with $RM_{z:C_{from}\rightarrow C_{to}}^{1}-RM_{z:C_{from}\rightarrow C_{to}}^{0}$

= - {[}$RM_{z:C_{to}\rightarrow C_{from}}^{1}-RM_{z:C_{to}\rightarrow C_{from}}^{0}${]},
which is compatible with the reassignment properties in  \ref{RM}.
This table allows simplifying reassignment computations in addition to providing a good means for studying the semantic influence of reassignment. Without pursuing this discussion any further in the present article, suffice it to say that these properties synthetically show that a vertex is inclined to either follow its neighbor or leave a community where a vertex has arrived to which it has no connection.

%\section{Reassignment and Nash equilibrium}
\section{Nash Equilibrium}\label{NE}

This section will demonstrate that applying reassignment leads to a Nash Equilibrium (NE), i.e. a situation in which no vertex has an incentive to leave the community where it has been assigned. To achieve this goal, we can interpret the vertex reassignment problem as a game theory problem. The  $n$ vertices may indeed be considered as $n$ agents seeking to optimize their benefit by attempting several strategies. Let's define a game of a finite set of players $n$, with each player $i$ having 
access to a finite set $S_{i}$ of strategies: $S_{i}={s_{i_{1}},s_{i_{2}}...,s_{i{}_{mi}}}$.
A strategy is a move executed by a player in order to derive some benefit.

Formally a game  $G = (S, f)$ where $S=S_{1}\times S_{2}...\times S_{n}$
is the set of strategy profiles. 
A strategy profile (or strategy vector) $s\in S$
is the combination of strategies of all players at a particular time of the game,
with each player applying just one strategy. Each strategy profile corresponds to a gain function

$f_{i}:S\longrightarrow\mathbb{R}$ for the player $i$ and $f={(f_{1},f_{2},...,f_{i},...,f_{n})}$.

It is important to note that player $i$'s gain depends at a particular point in time on the strategies adopted by the entire set of players. 
A strategy profile $s*$ is a Nash Equilibrium (NE) if no player is able to unilaterally change his/her own strategy
without causing a strategy change for one or more other players.
Formally speaking, a strategy profile is a NE if for all players $i$ and for all alternative strategies
 
 $s'_{i}\in S_{i}$, $s'_{i}\neq s_{i}^{*}$
, $f_{i}(s_{i}^{*},s_{-i}^{*})\geq f_{i}(s_{i}',s_{-i}^{*})$ where $s_{-i}$
is a strategy vector once the agent $i$'s strategy has been removed.

Three questions arise: 

(i) what are the conditions where an NE actually exists, 

(ii) if an NE exists, how can it be reached,

(iii) can this NE be reached in a polynomial time (in fact, the Np Completeness problem is more complex).

For a more detailed presentation, the interested reader is referred to (Papadimitriou,
book \cite{Nisan2007})). The existence of an NE remains a problem for any pure strategy game, in which players proceed deterministically. In contrast, a mixed strategy game is played using a probabilistic function with respect to the decision choices of all players for their next strategy. The Nash theorem asserts that in a game with  $n$ players ($n$ being finite) using mixed strategies,
at least one NE exists. 

This theorem is valid for mixed strategies but may not be directly applicable to pure strategies. It is possible however to provide an answer to the existence of a NE and the possibility of convergence towards this equilibrium in the case of a finite pure strategy game, should we be able to define a potential function that allows reaching a global optimum by seeking local optima for all agents. Moreover, if the search for a local optimum can be conducted in a polynomial time, then the problem enters into the PLS Complete problem class  ((Polynomial
Local Search) (Eva Tardos \& Tom Wexler \cite{Nisan2007}).
For any finite game, an exact potential function $\Phi$ associates a real value $\Phi(s)$ to each strategy vector $s$ 
under the following conditions:
\begin{eqnarray}
\Phi(s')-\Phi(s)=f_{i}(s')-f_{i}(s)
\end{eqnarray}
where $s=(s_{i},s_{-i})$ and $s'=(s'_{i},s_{-i})$

This function is interpreted as follows:

if player $i$ shifts unilaterally from strategy $s_{i}$ to strategy $s'_{i}$
with a gain of $f_{i}(s')-f_{i}(s)$ then the potential function increase with the same value
for all players; all players are then favorable to $i$'s change in strategy.
With this potential function, the Nash Equilibrium definitely exists, and we may converge to it by each player searching for local optima. In the community detection case, we are in a pure strategy game where agents are the $n$ vertices and strategies are the communities where these vertices are seeking to be attached. To ensure convergence to a Nash Equilibrium, i.e. a community partition scheme that satisfies all players, we must build a potential function $\Phi(s)$. 
The function $RM_{w:C_{1}\rightarrow C_{2}}$
which represents the expected gain for vertex w once it has been reassigned, takes into account the modularity gain of all communities and all vertices. In other words, $\Phi(s)=f_{i}(s)$.
This choice of a local gain ensures that the local vertex reassignment algorithm detailed in the previous section converges towards a Nash Equilibrium. Since the 
$RM$ computation is polynomial, the convergence algorithm towards the Nash Equilibrium is PLS Complete. With this result now confirmed, let's note that the NE may not be unique and, furthermore, the NE reached is not necessarily the best one.

\section{Experiments\label{sec5}}

In \cite{Crampes2013}, we have shown the possibility of analyzing unipartite, bipartite, and directed graphs indifferently, which has led us to conduct two types of experiments. On the one hand, we applied our reassignment algorithm to traditional unipartite graphs like 'Karate Club' of 'Dolphins' to compare our results with those of other authors. On the other hand, we applied our algorithm to community detection in bipartite graphs.

\subsection{Unipartite Graphs}

 \subsubsection{'Karate Club' benchmark \label{sub:karate}}

\begin{figure}
\includegraphics[scale=0.3,bb=0 0 800 800]{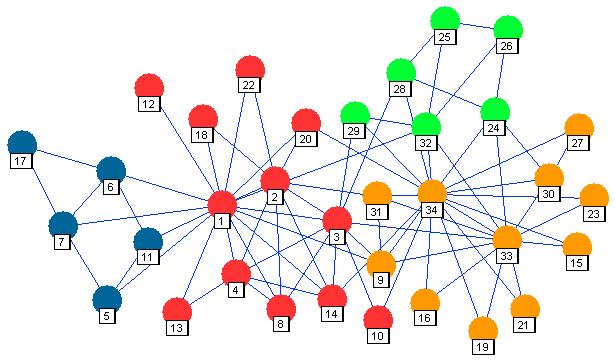}\caption{"`Karate Club"' graph with partitioned communities}
\label{karate}
\end{figure}

Applying the Louvain algorithm to the 'Karate' graph from Zachary \cite{Zachary1977}], as displayed in Figure \ref{karate}, outputs four communities. 

Our reassignment measure produces no instability since all RM values are negative. The computed modularity is higher than that produced by all other known algorithms. In particular, it is much higher (0.470) than the one announced by the authors using a local approach to reach a Nash Equilibrium, i.e. 
\cite{RNarayanam2012} and \cite{Chen2011} which announce 3 communities.

\begin{figure}
\includegraphics[width=4.7cm,bb=0 0 800 800]{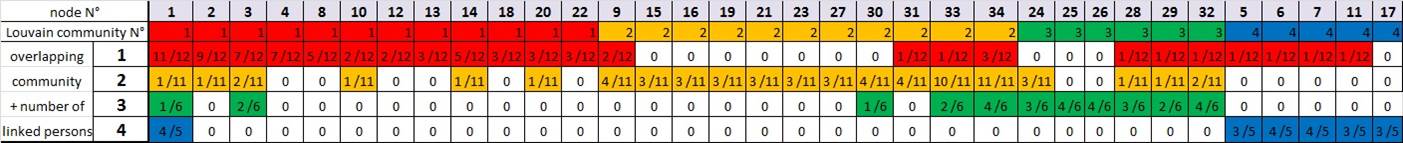}\caption{"Karate Club" Nash Equilibrium}
\label{karate-NE}
\end{figure}

\subsubsection{"Dolphins" benchmark \label{sub:dolphin}}

With respect to the 'Bottlenose Dolphins' graph, a social network of dolphins living in Doubtful Sound (New Zealand) 
 \cite{Lusseau2003} (number of vertices: 62, number of edges: 159), Louvain produces 4 communities with a modularity equal to 0.48; our reassignment measure indicates 4 unstable vertices. After reassigning these vertices and after reaching the associated Nash Equilibrium (see Figure \ref {karate-NE}), the modularity results rises to 0.51. Other authors have found a lower modularity value, except for \cite{RNarayanam2012}, whose results however could not be verified due to a lack of appropriate data. Without any further analysis, our approach interestingly seems to improve on the Louvain method, with above all a polynomial means for verifying stability. Compared to other methods using Nash Equilibrium, we have produced better results or, in some cases, equally good results (should these results be verifiable).

\subsection{Bipartite Graphs} \label{sub:SW}

 \subsubsection{'Southern Women'  benchmark\label{sub:Southern-Women}}

This benchmark has been studied by most authors interested in checking their partitioning algorithm for bipartite graphs. The goal here is to partition, into various groups, 18 women who attended 14 social events according to their level of participation at these events. In his well-known cross-sectional study, Freeman \cite{Freeman} compared results from 21 authors, most of whom identified two groups.

\begin{figure}
\includegraphics[width=8.3cm,height=5cm]{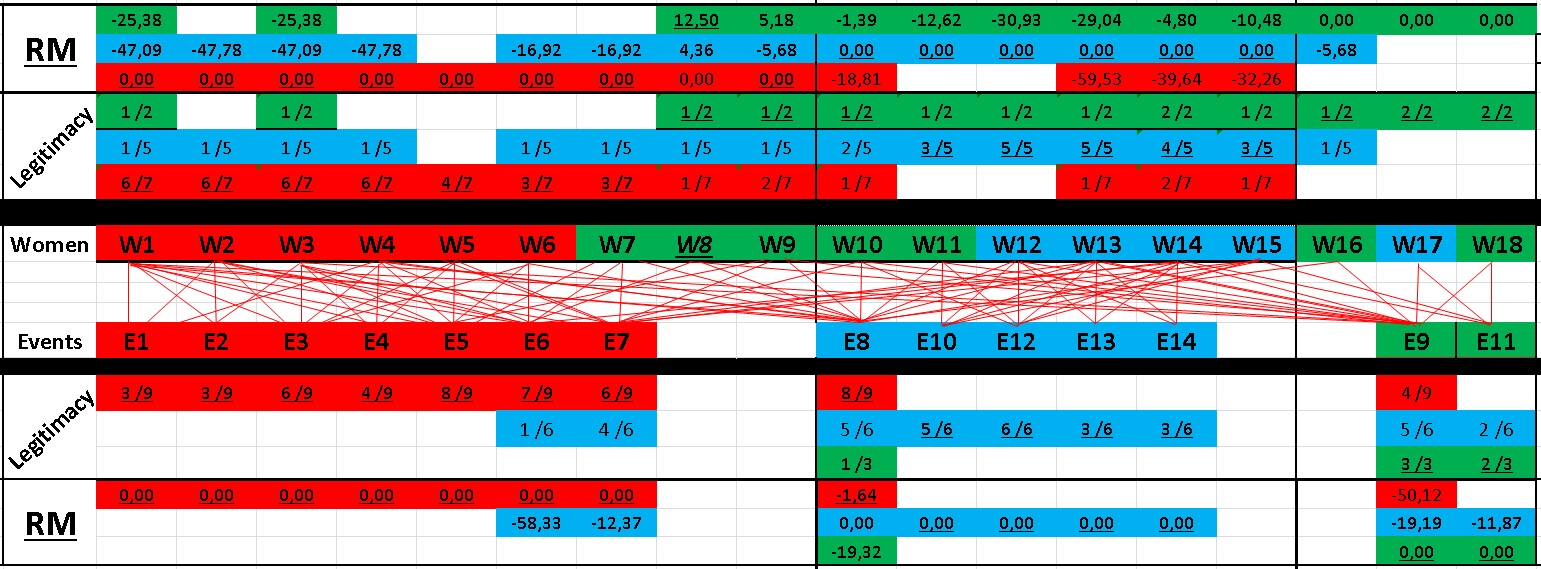} 
 \caption{Bipartite Graph Women Events: Communities, modularity, legitimacy and RM measures}

\label{WE} 
\end{figure}

\emph{Results. }In Figure \ref{WE}, the bipartite graph is depicted as a bi-layer graph in the middle with women at the top and events below; moreover, the edges between women and events represent woman-event participations. Three clusters with associated women and events have been found and eventually highlighted in red, blue and yellow coloring. This result is more accurate than the majority of results presented in \cite{Freeman}; only one author found three women communities. Beyond mere partitioning, Figure \ref{WE} presents overlapping communities using two overlapping functions, namely legitimacy and reassignment modularity (RM). Legitimacy and RM for women are placed just above women partitioning; for the studied events, both are symmetrically shown below event partitioning. As expected, reassignment in the same community produces a zero RM value. The best values for legitimacy and RM have been underscored. A positive value of RM indicates that the corresponding vertex could have been a member of another community; this is the outcome of an early assignment during the first Louvain phase for entities with equal or nearly equal probabilities across several communities. It can be observed in \cite{Freeman} that woman 8's community is also debated by several authors; our results appear to be particularly pertinent in terms of both partitioning and overlapping.

\begin{figure}
\includegraphics[width=5.3cm,bb=0 0 800 800]{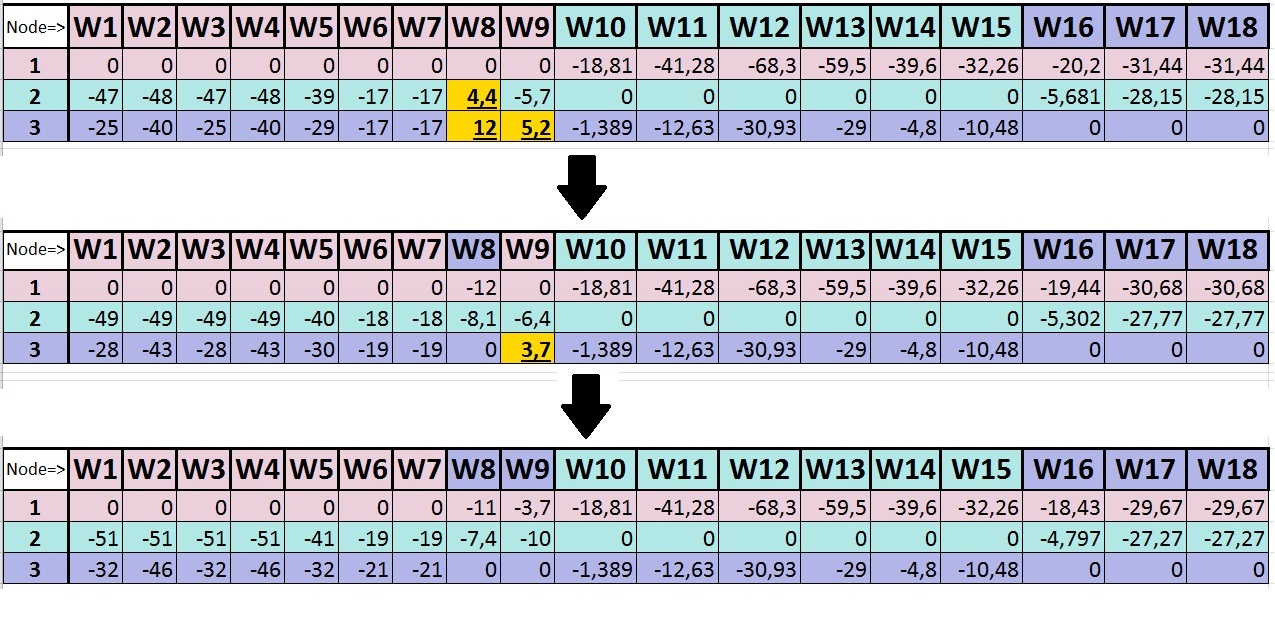} 
%\vspace{-0.2cm}
 \caption{WE: Achieving Nash Equilibrium}
\label{WE2} 
\end{figure}

The fact that women and events are correlated may be considered to cause a bias, such as in the number of communities. When comparing our results to those of other authors however, the merging of our blue and green communities produces their corresponding second community. In a trial designed to obtain a varying number of communities in both sets, Suzuki \cite{Suzuki2009} found a large number of singletons. These results were far from those presented in \cite{Freeman}, while ours were compatible and more highly detailed.

We complete these results with both overlapping and Nash Equilibrium. Figure \ref{WE2} shows how the Nash Equilibrium is reached by a three-step process, with each step displayed as a horizontal table.

This outcome is more precise than the majority of results presented in  
 \cite{Freeman}. measures show with greater accuracy this tendency to prefer association with another community. More specifically, persons W8 and W9 are unstable. After reassignment, the three communities all become stable. Overall modularity rises from 0.309 to 0.325. It is interesting to note that the two reassigned vertices become members of the third community, which is strengthened, while no other author even detected this community.

In each table, the top row lists the vertex figures and their assigned community (a different color for each community). The first column on the left gives the community number. During the first phase, we applied the 'Louvain' algorithm. The numbers in each cell indicate the RM value for each vertex (column) if placed in a community (row). We can observe three unstable situations in the first horizontal table. During the second phase, as observed in the second horizontal table, we applied our reassignment algorithm based on RM measures. The situation improves and modularity increases. One unstable situation still remains. During the third phase, seen in the third horizontal table, the situation has become stable and Nash Equilibrium has been reached: no one is willing to change community (all RM values turn negative).

 \subsubsection{Facebook account} 

This medium-sized experiment involves the Facebook account of a student, from which we have extracted photos and associated tags. The only impartial selection criterion was the number of tagged photos associated with the account. The dataset includes 644 photos for 274 different tags. These two types of information form a bipartite binary graph. A photo is considered to be connected to a tag (generally a person) if the photo has been tagged with this person's  name.
Photos  are  not linked  nor  are  the  persons related. 

\begin{figure}
\includegraphics[width=4.5cm,bb=0 0 800 1100]{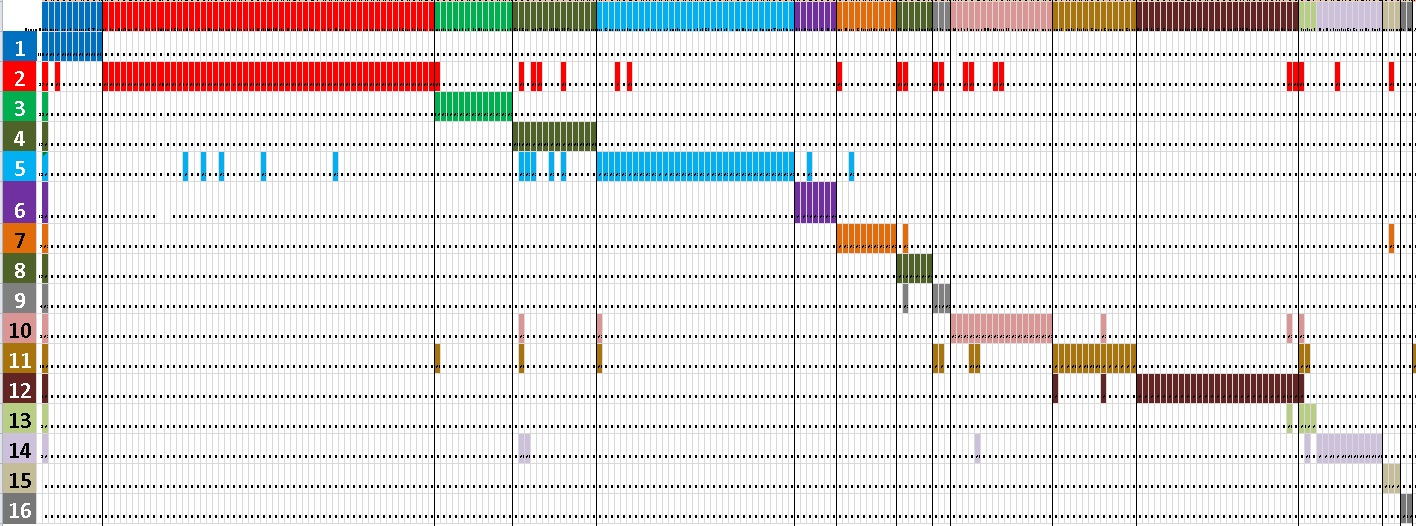}
\caption{Facebook account: Communities and overlapping information}
\label{facebook} 
\end{figure}

After cleaning the dataset, we applied to it our community detection method in order to obtain partition and overlapping information. We deleted the tags that were not persons (like 'landscape'). We initially obtained over 300 communities having one photo and one unique tag. These communities were the result of a single individual with just one associated photo and just one tag. We ignored these communities, but they may still be worthwhile for the owner. The other communities (i.e. with several persons) are presented in Figure \ref{facebook}, where the individuals are displayed as columns and the communities as rows. After a first pass with Louvain, 16 communities of more than 2 individuals were extracted: they overlapped only sparsely, which may be easily understood when considering that photos assume a physical presence of individuals at a particular time and place. They effectively expose several life periods of the account owner. We identified two types of individuals: those linked at a point in time during the owner's life (e.g. a classroom) were gathered in one community; and those (family members or close friends) present several times in his/her life may appear in several communities. The first community (blue) has the particular feature of containing an individual (first column on the left) included in all communities except 15 and 16. This person is the owner. Most photos for this community include the owner and others who are not present in other communities, which let's say then constitutes the owner's characteristic community. If we were only considering partitioning, then the owner would be a member of the one community in which he is isolated from other groups of friends in other communities. Thanks to overlapping information, we are able to see that the owner is also present in almost all communities. 

This experiment shows that for the studied dataset, if we were only to consider partitioning information, the results would present certain inconsistencies, like the owner's isolation in a particular community (other inconsistencies will not be discussed due to length limitations of our paper). Conversely, thanks to overlapping information and the association of both sets of individuals and properties in communities, a finer analysis could be performed. The RM measure only indicates one unstable person out of the 250, which underscores the proper initial set-up of communities found by Louvain. The $RM$ measure applied to photos results in 15 photos being unstable in their respective communities found by the Louvain partitioning set-up, which suggests that right from the beginning, the dataset has very few ambiguities.

On this dataset, Nash ffEquilibrium is reached in 18 steps. The modularity at the beginning just after Louvain is equal to 0.7431, and modularity at Nash Equilibrium amounts to 0.7491. This increase is quite noticeable. Since the Facebook matrix is quite large, it is not very useful to visualize each step. This visualization problem will be addressed in the next section.

\begin{figure}
\includegraphics[width=4.5cm,bb=0 0 800 1100]{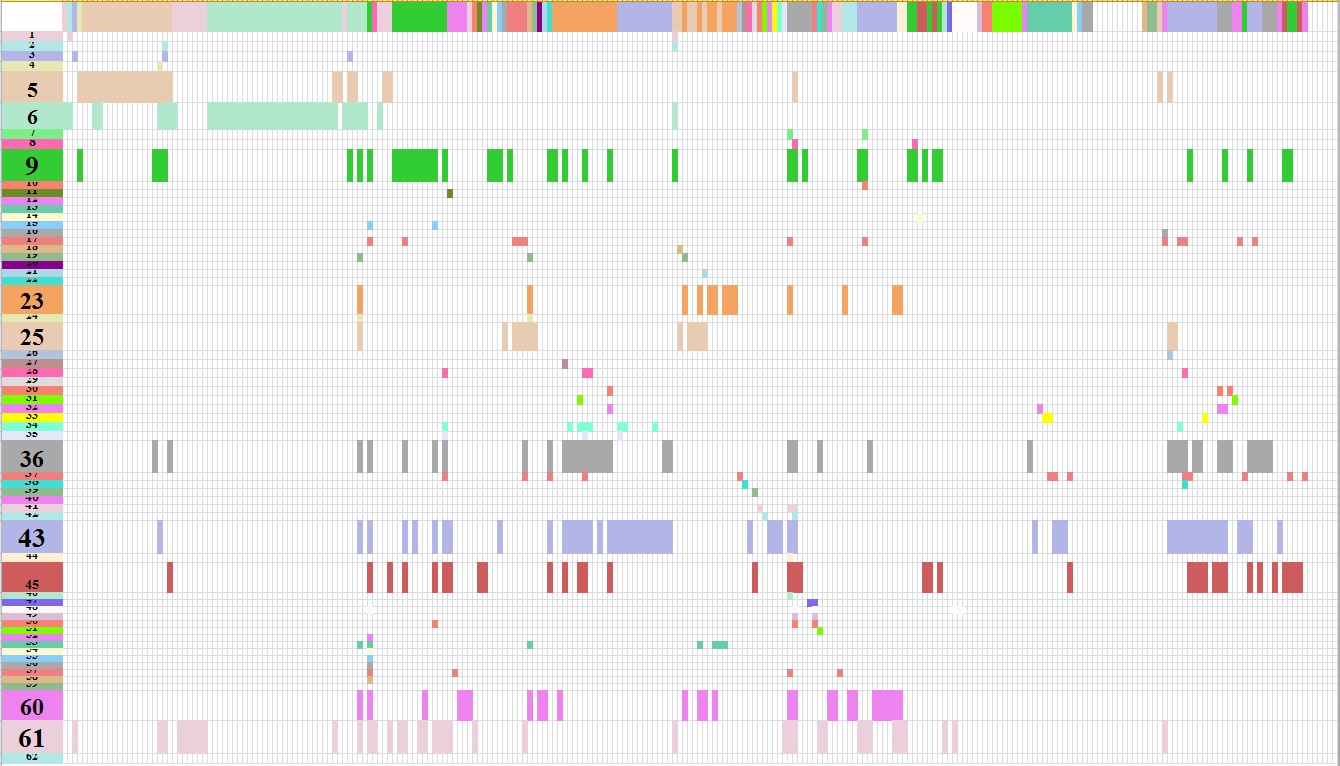}
\caption{New Facebook Account:\newline Before Nash Equilibrium}

\label{facebook2} 
\end{figure}

\begin{figure}
\includegraphics[width=4.5cm,bb=0 0 800 1100]{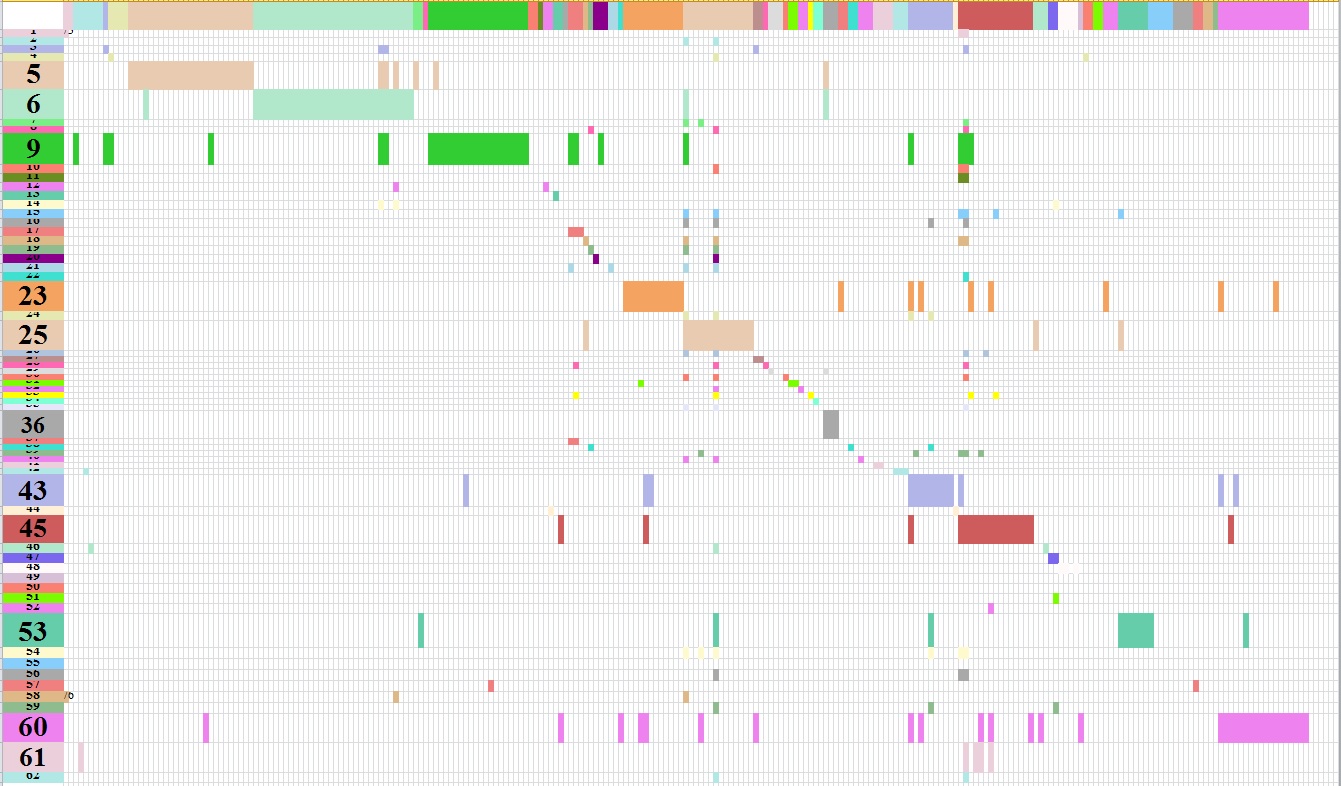}
\caption{New Facebook Account: \newline After Nash Equilibrium}
\label{facebook3} 
\end{figure}

On a new Facebook account quite of the same size, 
we computed the overlapping information from Louvain Algorithm
and then the Nash Equilibrium.
We show in Figures \ref{facebook2} and \ref{facebook3} 
The situation before equilibrium (right after Louvain Algorithm) and after reaching Nash Equilibrium.
The first line on each figure displays the partition situation with separate colors (i.e. each person
is represented as a colomn and each color corresponds to a community).

Nash equilibrium is reached after 1250 steps in less than one second.
We can see that communities after Nash Equilibrium are bigger and less scattered among persons.
There are 62 communities among which only 10 communities are of relative big size.
A lot of singletons are still present. There are in the dataset a lot of photos with only one person.
Moreover there are less overlapping potential situations after Equilibrium shown with legitimacy measure,
meaning that communities are more stables and persons are more confortable in their community.

The situation after Nash Equilibrium is much better for the whole graph and at the same time for each person.
Modularity increases a lot from Louvain to Nash Equilibrium from  a very low value showing some scattered community distribution
 to a very good value of 0.4748.

This experiment may be extended to generate results with industrial application. More specifically, this method may be used to automatically build photo albums and spread personally adapted albums to targeted individuals or to accomplish marketing goals or for personal dissemination.

\subsubsection{Known benchmarks and large graphs}

Our recent experiments have been conducted on several large graphs. On a unipartite graph with over 10,600 vertices cited in \cite{Chen2011},
Nash Equilibrium is reached after 80 iterations, with just a few seconds of computational time.

It is not common to find large bipartite graphs in the scientific literature. We evaluated our algorithms using medium-sized bipartite benchmarks found in \cite{Barber2007}.
One of them is a dataset describing cross-shareholdings (in order to avoid takeover by new shareholders) in Scotland at the beginning of the 20th century. The data describe 108 Scottish companies between 1904 and 1905, displayed by each company's type of activity, capital and shareholding group. The dataset has been reduced to Board Members holding the right to vote on multiple boards. Barber found around (sic) 20 communities, whereas our number was 15 with very interesting overlapping information among these communities. The global modularity after reassignment is equal to 0.71, while Barber found a lower value of 0.57. 

To evaluate the scalability of our method, we tested a much broader dataset of co-author relationships on scientific papers in order to extract scientific communities. 
The data were extracted from the well-known "Pubmed" library (http://www.ncbi.nlm.nih.gov/pubmed) of scientific papers in the biomedical field. The dataset contains 30,000 individuals and over 85,000 scientific papers. We extracted 184 communities of 670 members on average in less than 3 seconds with Louvain. The reassignment was then performed on more than 80,000 unstable vertices, demonstrating that the intermediary result from Louvain is quite unstable. 
The computing time of one hour and a half still requires some optimization.

Computations on large-scale data may be used to achieve industrial or administrative objectives. Communities form a workable knowledge base to pursue personal dissemination, targeted actions, dissemination-related savings, department restructuring according to community membership, etc.

\section{Discussion and \\ unresolved problems}

Regarding community detection problem in Social Networks, after applying methods inspired from data analysis, specific algorithms have recently appeared and prove to be very efficient in terms of computation time and when considering the results, although the solutions given are suboptimal. Several criteria for defining a good graph partition have been proposed, in particular modularity based on a sound probabilistic basis. More recently, entropy and Nash Equilibrium have yielded new and complementary results.

Some issues remain unresolved. It is obviously still possible to propose new heuristics that offer better results according to preexisting criteria (modularity, entropy, Nash Equilibrium) or that may be invented. On the other hand, many related difficulties need to be considered.

\paragraph{Visualization}

The most obvious and difficult problem seems to be visualization. Visualizing results is a prerequisite to providing interpretation. Displaying a unipartite graph partition is quite simple, as can be observed for the "karate" dataset in Figure 1. On larger graphs, if the number of communities remains rather low, the problem can still be solved, like in \cite{Blondel2008},
where the two main linguistic communities are prominently displayed, but the other communities with fewer linguistic properties are harder to distinguish. Regarding overlapping, visualization raises the challenging problem of hypergraph planarity, where solutions offering a clear visualization are difficult to find. For three communities, we may use Venn diagrams \cite{venn} for an effective visualization scheme, but with a higher number of communities Venn diagrams become very awkward to read. For this reason, we rely upon a matrix-like representation in this paper, though it does not display the initial graph form. When community overlapping needs to be shown on bipartite graphs, the loss of representation is even greater. For example, for Facebook datasets, we must narrow the display to just one type of vertex for the columns and communities for the rows. The other type of vertices must be displayed on another vertex-community matrix.

\paragraph{Constraints}

In many cases, we are seeking to extract communities in which some constraints may be added. One such constraint has already been discussed in the data analysis: the predefined number of communities. It then becomes possible to return to traditional methods like KNN or K-means that do not require modularity. With modularity, we may use a reassignment algorithm like the one presented in our paper. Until now however, we are not aware of any articles that study community restructuring, i.e. given an optimal detection and assignment, proceeding by eliminating or adding one or more communities. It is then essential to find a new optimal assignment on the new structure that takes into account the previous situation (if this previous situation is not taken into account, it is easier to consider the problem from the beginning and reassign every vertex).

Other problems are also of great interest. We have already discussed introducing another quantitative criteria like entropy or Nash Equilibrium, but more complex constraints may arise. For example, perhaps an equal number of individuals are sought for each community (let's recall here that individuals are related and the problem is not limited to dividing the population into equal parts; we must also optimize modularity or any other similar criteria). Other constraints may be considered, like banning two persons from membership in the same community. This problem has been extensively addressed in its simplest form, i.e. graph coloring. It takes on an additional dimension here because we are also seeking to optimize other criteria. Let's note that some problems have already been studied within the hypergraph domain \cite{35}.

Many other constraints are also possible. Let's mention in particular community detection with complementary individuals in bipartite graphs. Examples would include obtaining a unique property profile distribution in one or more communities or else an even distribution among communities. A simple example of this situation is a rugby team composed of players with complementary skills for the game. This same situation may be observed for a project team within a company.

\paragraph{Evolution}

Community evolution is the subject of a quite extensive body of research, e.g. \cite{Roth2008a}.
This field is especially interesting since it highlights social or ecological changes. The subject is even more appealing if we take into account the temporal evolution of a network. It is easy to understand the importance of some topics related to the possible causes of network and community evolution. For example, what is the expected time evolution for networks like Facebook or Twitter; what are the evolution scenarios dependent on new service offerings? This issue may be extended to the political arena when considering electorate community partitioning, which may depend on voting intention overlapping or networks influencing voters. In this case, game theory and Nash Equilibrium may still lead to new avenues if we design less consensual potential functions than those proposed in 
Section \ref{NE}.

\section{Conclusion  and perpectives}

%\vspace{-0.2cm}
 We have applied a community detection method for all types of graphs (unipartite, bipartite and directed). Our initial contribution was to detect and display overlaps. The second contribution was to detect instabilities produced by the detection algorithms in spite of their quality. To identify a stable solution, we designed a reassignment function and showed that this solution may lead to a Nash Equilibrium in a polynomial time. More generally, almost all community detection methods produce unstable membership situations. Our contribution allows complementing any community detection algorithm in order to find a stable equilibrium with optimal modularity in a polynomial time. It is important to notice the 'social' feature of all vertices, which combine their own interest with the collective goals. This point will be further developed in our subsequent work. Our future research will enhance the reassignment methods exposed herein. More specifically, one limitation of the present method is the fact that it takes into account the number of communities found by the initial partitioning algorithm. We will determine how to eliminate or add communities. The focus will then be on studying how adding or deleting policies can give rise to unstable or unsatisfactory situations. Generally speaking, while creating initial graphs and then detecting communities have been widely studied, only a few contributions have been provided on the application of external constraints and on adaption dynamics within communities. These models would be suitable for problems like adding or deleting production units, services, etc. In reality, graphs are often multipartite, which is why our approach mixes unipartite and bipartite graphs. We should also consider more selfish potential functions operating on less altruistic agents.

\bibliographystyle{abbrv}
 \bibliography{communitiesv2}

\end{document}